\documentclass{article}

\usepackage{arxiv}

\PassOptionsToPackage{numbers,sort&compress}{natbib}

\usepackage[utf8]{inputenc} 
\usepackage[T1]{fontenc}    
\usepackage{hyperref}       
\usepackage{url}            
\usepackage{booktabs}       
\usepackage{amsmath}        
\usepackage{amssymb}        
\usepackage{amsfonts}       
\usepackage{nicefrac}       
\usepackage{microtype}      
\usepackage{graphicx}
\usepackage{natbib}
\usepackage{doi}

\title{TOMOYO Linux: A Mandatory Access Control Method Based on Application Execution State}

\author{{Toshiharu Harada}\\
	Graduate School of Information Security\\
	Institute of Information Security\\
	Yokohama, Japan\\
 	\And
    {Tetsuo Handa}\\
    Solution Services Division \\
	NTT DATA INTELLILINK Corporation\\
	Tokyo, Japan\\
    \And	
    \href{https://orcid.org/0000-0001-5596-282X}{\includegraphics[scale=0.06]{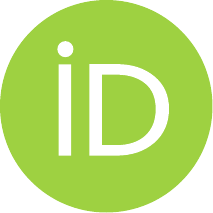}\hspace{1mm}Masaki Hashimoto}\\
	Graduate School of Information Security\\
	Institute of Information Security\\
	Yokohama, Japan\\
 	\And
	{Hidehiko Tanaka}\\
	Graduate School of Information Security\\
	Institute of Information Security\\
	Yokohama, Japan\\
}

\date{}

\renewcommand{\shorttitle}{TOMOYO Linux: MAC Based on Application Execution State}

\hypersetup{
pdftitle={TOMOYO Linux: A Mandatory Access Control Method Based on Application Execution State},
pdfsubject={cs.CR},
pdfauthor={Toshiharu Harada, Tetsuo Handa, Masaki Hashimoto, Hidehiko Tanaka},
pdfkeywords={Access control, Mandatory access control, TOMOYO Linux, Program execution history, Domain, Operating system security},
}

\begin{document}
\maketitle

\let\thefootnote\relax\footnotetext{
	This document is an English translation of the authors' paper originally published in Japanese: T.~Harada, T.~Handa, M.~Hashimoto, and H.~Tanaka, ``A Mandatory Access Control Method Based on Application Execution State,'' IPSJ Journal, Vol.~53, No.~9, pp.~2130--2147 (Sep.~2012).
	\copyright~Information Processing Society of Japan (IPSJ).
}

\begin{abstract}
Existing access control methods grant access requests based on the combinations of applications as subject and files as objects. Therefore intents of applications and the possible effects caused by granting the access requests have not been taken into consideration. In this paper, we propose a new access control method based on application history and intents. With our access control method, system administrators can reduce the risks caused by malicious access attempts and wrong operations. In this paper, the concept and implementation design will be explained as well as the brief evaluation report of TOMOYO Linux, our implementation of the new access control method to Linux.
\end{abstract}

\section{Introduction}
In recent years, the scope of information systems has expanded and the processing they perform has become more sophisticated, so the software that realizes them has grown in scale and complexity. Ensuring information security has therefore become a challenge. For example, if the access permissions of files and directories are not configured appropriately, leakage of information or system malfunction can easily occur; and even if they are configured appropriately, there remains a latent risk that the information system will be taken over through unauthorized access, worms, and the like. These risks cannot be eliminated completely, and as software grows larger and application processing grows more complex, they will surely increase in the future.

Sandboxing~\cite{Peterson2002}\cite{Oyama2003} and access control are two techniques for ensuring information security. A sandbox runs an application in a restricted environment: by running an untrusted application, such as a file downloaded from a network, inside an environment isolated from the rest (the sandbox), the application can be confined~\cite{Goldberg1996} and the scope of damage limited. Access control is an elemental technology for realizing a sandbox: for the access requests that arise as an application runs, it decides whether each request may be carried out. An information system functions through applications requesting operations such as reading and writing files, executing other applications, and network communication, with the kernel processing those requests. Therefore, by appropriately controlling the access requests of individual applications, it is possible to prevent uses of the information system that the system administrator does not desire.

Conventional access control methods decide whether to permit access based on the combination of an application and the target file it attempts to access. The execution state of the requesting application is not taken into account, and because it cannot be used as a condition (decision criterion) of access control, situations that compromise information security have arisen. For example, an application that performs a wide variety of processing, such as a web browser (an HTTP client), should essentially change the access it is granted according to its execution state~\cite{barth2008security}; but conventional access control methods cannot express an execution decision that takes this into account. As a result, sandboxes based on conventional access control methods are coarse-grained, and the localization and confinement of damage have been insufficient.

Therefore, in this paper, regarding access control as a fundamental technology for ensuring information security, we propose a new access control method capable of taking the execution state of an application into account. In the proposed method, the execution state of an application is interpreted from the history leading from system startup up to the execution of the application in question, together with information such as the application's command-line arguments and the environment variables at the time the request occurs; access is decided using this information as conditions. In addition, because the proposed method performs the acquisition, decision, and enforcement of information from within the kernel~\cite{Loscocco1989}, access control can be enforced safely and without omission.

The remainder of this paper is organized as follows. Chapter~2 reviews conventional access control methods and describes their problems. Chapter~3 describes the basic idea for realizing the proposed method. Chapter~4 takes up TOMOYO Linux, which the authors have been developing, as an implementation of the proposed method for the Linux operating system (hereafter OS), and uses example policy descriptions to show the access control the proposed method makes possible. Chapter~5 describes, as an evaluation of the method, the features of the method, the results of an initial study of its usability, and its impact on performance. Chapter~6 discusses a comparison of the proposed method with other methods, its resistance to unauthorized access, and remaining issues. Finally, Chapter~7 summarizes the proposal and presents conclusions.

\section{Limits and Issues of Conventional Access Control Methods}
This chapter first reviews existing research, centering on the two access control methods used in most of today's operating systems, and organizes the basic terminology related to access control used in this paper. It then describes the limits of the conventional methods and organizes the issues to be solved.

\subsection{Conventional Access Control Methods}
The access control methods used in most information systems today can be broadly divided into two: Discretionary Access Control (DAC) and Mandatory Access Control (MAC)~\cite{Bishop}. DAC lets the owner of a file or other resource control which users may access it; because it uses identity information to decide whether to permit access, it is also called identity-based access control. MAC, on the other hand, does not allow even the owner to control which users may access a resource: each access decision is made according to rules defined by the system administrator. Because MAC requires rules for deciding access, it is also called rule-based access control.

MAC requires rules that serve as the criterion for access decisions, and these are generally called a policy. A policy requires some identifier for the subjects and objects that are the targets of access control. In the label-based MAC (Labeled Security) defined by the TCSEC (Trusted Computer System Evaluation Criteria)~\cite{tcsec1983}, published by the U.S. Department of Defense in 1983, this identifier is called a ``label.'' After the TCSEC, the only MAC implementations were label-based; but since 2006, mandatory access control using path names as identifiers has been proposed, and these are called pathname-based MAC.\footnote{\url{http://lwn.net/Articles/277833/}} For example, the standard source code of the Linux kernel currently includes four MAC implementations: SELinux~\cite{PeterLoscocco2001}\cite{Loscocco2001}\cite{Smalley2005}, SMACK~\cite{Schaufler2008}, TOMOYO Linux, and AppArmor~\cite{Cowan2000a}. Of these, SELinux and SMACK are label-based MAC, while TOMOYO Linux and AppArmor are pathname-based MAC.

A policy is defined as a collection of conditions describing what is and is not permitted. The component that interprets a policy and realizes it is called the policy mechanism. In discussions of access control, the terms subject and object are generally used, and this paper follows that convention. So far we have used ``application'' loosely to refer to computer programs in general, but before entering the main discussion we make the definition precise. Hereafter, ``application'' in this paper refers to an executable file recognized by the OS process-execution instruction (the \verb|execve| system call on Linux). The unit of execution, or instance, of an application is called a process (OS process).

Whereas the various MACs above are applied automatically (passively) within the kernel, there are also methods that perform access control based on the application's own declaration: seccomp~\cite{Winter2008} on Linux and Capsicum~\cite{watson2010capsicum} on FreeBSD. With seccomp, once a process itself issues the system call \verb|prctl(PR_SET_SECCOMP, 1);|, it can thereafter execute no system calls other than the four \verb|read()|, \verb|write()|, \verb|exit()|, and \verb|sigreturn()|. Current seccomp can be applied only to special uses such as grids, but a proposal has been made to generalize the usable system calls. Capsicum makes it possible to restrict the capabilities of a process per application and, like seccomp, is based on the process's own declaration. Because Capsicum cannot be applied to uses such as controlling which files each application may open, it cannot by itself directly solve the problem addressed in this research.

\subsection{Limits of Conventional Access Control Methods}\label{sec:genkai}
DAC is widely used as the most basic access control method. Because DAC decides access based on user information, the access permissions it can define are coarse---read, write, and execute---and, moreover, users with administrative privileges are not affected, which is a limitation. For this reason, if control is seized through a buffer overflow attack~\cite{Ken2002-01-15} or the like, the effect of limiting the damage is weak; in particular, if control of a process running with administrative privileges is seized, serious damage cannot be avoided.

Label-based MAC can compensate for the limits of DAC. In label-based MAC, access is decided based on the labels assigned to subjects and objects. By defining labels for subjects and objects as needed, access control at an arbitrary granularity becomes possible, and processes running with administrative privileges are not treated as exceptions but are also subject to control. However, even if label-based MAC can appropriately control whether a subject may access an object, that alone does not mean information security can be maintained. We give two examples.

As a first example, the Apache web server, if a file named \verb|.htaccess| exists in the directory where content is stored, interprets the contents of that file as configuration and does not treat it as web content (does not send it to the client). However, if that file is renamed to \verb|index.txt|, its contents will be delivered to the client as content. To prevent this using label-based MAC, the configuration file \verb|.htaccess| and content files such as \verb|index.txt| would have to be assigned different labels, and the Apache web server process would have to be implemented so that, after opening a file, it obtains the label of the opened file and decides how to treat it according to the label's contents. In reality, however, there is no common convention or agreement about which labels to assign to which files,\footnote{On Fedora~15, when a \texttt{.htaccess} file is created under /var/www/html, it is assigned the same label as content.} and most applications are implemented to decide their behavior and handling based on file names.

As a second example, the server of the SSH (Secure Shell) service has an option to display the contents of a specified file as a banner when accepting a request from a client. For instance, running it as \texttt{/usr/sbin/sshd -o 'Banner /etc/shadow'} would disclose the passwords stored in \verb|/etc/shadow| to a client before authentication. The first example shows a case where a file name affects an application's processing; the second shows a case where the command-line arguments given to an application determine its processing. Neither can be handled by conventional access control.

Label-based MAC does not use names (file names) as identifiers. The chief reason is that a name is an attribute of the subject or object and can change. By assigning labels to entities, access control that is unaffected by name operations such as renaming becomes possible, and leakage of object information can be prevented; but even if labels are maintained and the access conditions on objects are preserved, that alone is not sufficient.

The two examples above concern the processing of the applications \verb|Apache| and \verb|sshd|, respectively, and arise as effects of having permitted the access of renaming a file and of executing an application. Because conventional access control does not take these into account, it cannot restrict them.

\subsection{Issues to Be Solved}
\label{sec:purpose}
Conventional access control methods decide access based on the combination of subject and object, and do not consider what kind of processing the subject application is about to perform. In the first example shown in Section~\ref{sec:genkai}, the problem is that the rename request is permitted without considering how the information stored in the object will come to be used; in the second example, the problem is that execution of the SSH server is permitted without considering that the contents of the password file will be displayed to a client before authentication. In other words, conventional access control methods have the following issues.
\begin{itemize}
\item In deciding whether to permit an access request, how the information stored in the object will be used is not considered.
\item In deciding whether to permit an access request, what effect will arise on the operation of the information system or the application is not considered.
\end{itemize}

The above issues can be solved by incorporating the following mechanisms into the access control mechanism.

\def\theenumi{\roman{enumi}}
\begin{enumerate}
\item A mechanism to acquire (classify), for each application, the state in which it is being executed.\\
Even for the same application, the access permissions it requires differ according to the state in which it is executed. Therefore, it is necessary to use some criterion to classify the state in which an application is executed, and to perform access control per classification.
\item A mechanism to judge, for an application, what it is about to do.\\
Even if classification by state is achieved, a criterion is needed to infer the intent of each access request issued by each application and to judge the effect of permitting the access.
\item A mechanism to make access decisions for all applications that take their state and intended processing into account.\\
An access control mechanism based on (i) and (ii) above must be implemented so that it cannot be bypassed and applies without omission.
\end{enumerate}

\section{Proposal of an Access Control Method Based on Execution State}\label{sec:Teian}
This chapter describes, as the basic ideas for realizing a new access control that compensates for the limits of conventional access control described in Chapter~2, the classification of applications using execution history and the access decision based on their processing.

\subsection{Classifying Applications Using Execution History}
To perform access control that takes each application's processing into account, the access control mechanism must be aware of the application. Even for the same application, its handling needs to change according to the state in which it is executed. For example, a shell (the command interpreter in Linux) obtained by logging in from the server console, a shell obtained by logging in from an SSH client, and a shell that the Apache web server obtains in order to execute a CGI are identical as program files, but the access permissions they should be granted should be treated separately. Moreover, the access control addressed in this paper is meant to maintain information security, and it is desirable to cover all applications. That is, for every application it must be possible to separate cases according to its execution state. In this paper, we propose a method that classifies the execution state of an application by managing the execution history of each application.

Here, the execution history is, for each process of an application, a list of the program names of the applications that led to its execution. Table~\ref{tab:pih} shows examples on Fedora~15, one of the Linux distributions.\footnote{These results differ depending on the distribution and the execution environment.} All three examples concern processes of \verb|/bin/bash|, a shell. The \verb|/sbin/init| at the head of items 1 through 3 is the process created at Linux startup, but the history by which \verb|/bin/bash| came to be executed from there differs among the three. Item~1 shows that \verb|sshd|, a daemon executed from the service startup script \verb|/etc/rc.d/init.d/sshd|, created a new process in response to an access request from a client, and \verb|/bin/bash| was executed from there. Item~2 shows that \verb|/sbin/agetty|, the application that accepts logins, executed \verb|/bin/login|, the application that performs login processing, and \verb|/bin/bash| was executed from there. Item~3 shows that the \verb|/bin/bash| executed as item~2 accepted \verb|su| (switch user), ``the application for executing commands as another user,'' and as a result \verb|/bin/bash| was executed.

In its standard state, the Linux kernel cannot determine, for the three example \verb|/bin/bash| processes, under what circumstances each was executed. Therefore it cannot distinguish them by the situations in which they are placed, nor define access permissions appropriate to each process's situation. However, if the contents of each process's execution history can be referenced, the access permissions granted to each can be distinguished---for example, allowing all applications to be executed without restriction when logging in from the console, while restricting execution to only specific applications when logging in from an SSH client.

\begin{table}[t]
\caption{Examples of Program Execution History (Linux).}
\label{tab:pih}
\centering
{\footnotesize
\begin{tabular}{|p{0.04\linewidth}|p{0.86\linewidth}|}\hline
1& bash process executed after logging in from the SSH service\\
 & \verb|/sbin/init /etc/rc.d/init.d/sshd /usr/sbin/sshd /usr/sbin/sshd /bin/bash| \\\hline
2 & bash process obtained by logging in from the console \\
 & \verb|/sbin/init /sbin/agetty /bin/login /bin/bash| \\\hline
3 & bash process obtained by logging in from the console and running the \texttt{su} command to work as administrator \\
 & \verb|/sbin/init /sbin/agetty /bin/login /bin/bash /bin/su /bin/bash| \\\hline
\end{tabular}
}
\end{table}

\subsection{Access Decision Based on Application Processing}
An application's processing is defined by the application's author and recorded inside the application's file. Therefore, to fully grasp an application's processing, the best approach would be to have the application itself declare it; but at present there is no criterion or description method applicable to all applications. Also, considering the possibility that an application is hijacked, there is no problem in reducing privileges, but accepting a declaration that increases privileges would create a new risk. Even if these problems were solved, modification of existing applications would still be required. Against this background, judging an application's processing by a direct method is difficult; but by using indirect methods, it is possible, within a certain range, to judge an application's processing without requiring modification of the application.

There exist factors that affect an application's processing. If these can be used as conditions (parameters) when granting access permission, inappropriate requests can be identified and rejected. We give some concrete examples of factors that affect an application's processing.

\begin{itemize}
\item Command-line arguments (many applications accept option specifications)
\item Environment variables (the command search path referenced at execution, proxy servers, etc.)
\item User input (processing branches according to user instructions)
\item Input data (the contents of data are analyzed and the result displayed)
\item Configuration files (many applications have their own configuration files and reference them at startup)
\item Libraries (there exist attacks that deliberately cause a different library to be referenced)
\item Time
\item The executing user (e.g., performing processing only for the administrator)
\item The presence or contents of files with specific names (\verb|/etc/nologin|, \verb|.htaccess|, etc.)
\end{itemize}

It is worth noting that the information above is already managed within the kernel, so it can be referenced as a parameter within the processing of MAC. In this paper, we propose using, within MAC, the information that affects an application's processing as condition/criterion parameters for access decisions in the policy.

\section{Implementation of the Proposed Method on Linux}\label{sec:tomoyo}
This chapter explains an implementation of the access control method that takes application processing into account, using TOMOYO Linux~\cite{Harada2009}\cite{Harada2010} as an example. TOMOYO Linux is both an example implementation of MAC on Linux and the name of the project that promotes it. Among the authors, Harada and Handa carried out the study of the method and functions of TOMOYO Linux, and Handa alone carried out the implementation of the proposed method in the Linux kernel and the coding of standard tools such as the policy editor. TOMOYO Linux has been extended from time to time in response to the study of the proposed method, and version~1.8.3, the latest at the time of writing, corresponds to the contents of this paper. The source code of TOMOYO Linux is published on SourceForge.jp.\footnote{\url{http://tomoyo.sourceforge.jp/}} Since mainline kernel version~2.6.30, a function-limited version of TOMOYO Linux, created for the mainline, has been included as a standard feature.

\subsection{Acquiring the Application Execution State}
In its standard state, the Linux kernel does not manage what history an executing application has. Each application process remembers the id of the parent process that created it, but because the information of terminated processes is removed from the kernel, the parent process can be traced back only within a limited range.

If, for every application process, we could give it the history leading from a common base point to the creation of that process, and retain it even after the process terminates, then the execution state of each application could be uniquely determined.

\subsubsection{Definition and Realization of the ``Program Execution History'' Concept}
In TOMOYO Linux, in order to make it possible to acquire the application execution state, the concept of program execution history was realized. The program execution history is, for each process, the path names of the applications that were executed up to the creation of that process and the program name (path name) of the running application, arranged with the single-byte space character as a separator. The contents of Table~\ref{tab:pih} shown earlier are, as they stand, examples of the program execution history concerning \verb|/bin/bash|.

\begin{figure}[t]
 \centering
  \includegraphics[width=.9\linewidth]{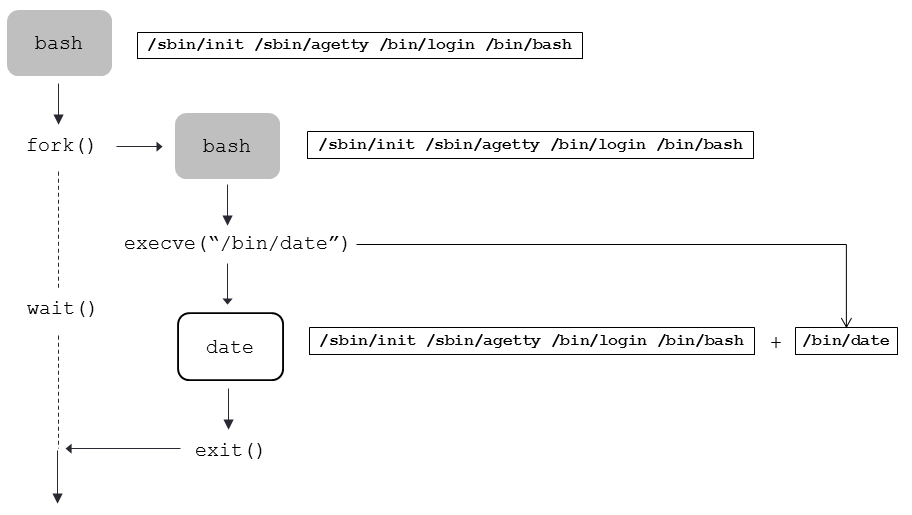}
 \caption{Defining Program Execution History.}
 \label{fig:fork}
\end{figure}

By using the mechanism of application execution in Linux, every process can be given its program execution history. We explain the method using a figure. Figure~\ref{fig:fork} shows the contents of the system calls (the interface by which an application invokes kernel functions) issued when \verb|/bin/date| (the command that displays and sets the date) is executed from a \verb|/bin/bash| obtained by logging in on Linux, and its result is awaited. In OSes that descend from UNIX, such as Linux, application execution is performed by a characteristic procedure. First, the (parent) process that is about to execute an application creates a process that is a copy of itself by the \verb|fork()| system call, and the copied process has its contents (process image) replaced by those of the application by invoking the \verb|execve()| system call. In Figure~\ref{fig:fork}, the \verb|/bin/bash| process issues the \verb|fork()| system call, and its copied process issues the \verb|execve()| system call, whereby \verb|/bin/date| is executed (when the \verb|/bin/date| process terminates, processing returns to the \verb|/bin/bash| process). Displayed to the right of each process is that process's program execution history. We see that its contents are inherited (copied) from the parent process at process creation, and can be updated by appending the program name of the application to be executed when \verb|execve()| is performed.

\subsubsection{Domains}
In TOMOYO Linux, access control is performed in units called domains. We describe domains in TOMOYO Linux. A domain is the program execution history described above with its base point added. For the state in which no application is being executed, namely ``the kernel itself (the kernel thread),'' we virtually denote its program name as \verb|<kernel>|. By definition, every domain necessarily begins with \verb|<kernel>|, so \verb|<kernel>| is a common base point. By introducing a base point, it becomes possible to distinguish a part of a program execution history from the whole.

Specifying a domain uniquely determines the contents of a program execution history. When the program execution history of a process matches a domain, the process is said to belong to that domain. \verb|/sbin/init|, the first process created in Linux, belongs to the domain \verb|<kernel> /sbin/init|. The domain to which a \verb|/bin/bash| obtained by logging in from the console belongs is \texttt{<kernel> /sbin/init /sbin/agetty /bin/login /bin/bash}.

A domain has the following properties.

\begin{itemize}
\item Every domain has exactly one parent domain.
\item Every process belongs to exactly one domain.
\item Even if the executing applications are identical, they do not necessarily belong to the same domain.
\end{itemize}

\begin{figure}[t]
 \centering
  \includegraphics[width=.9\linewidth]{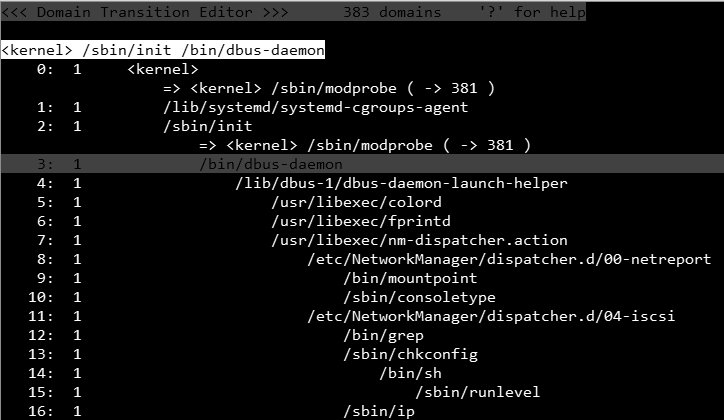}
 \caption{Domain Transition Example (Fedora 15).}
 \label{fig:fig-domain}
\end{figure}

In TOMOYO Linux, the program execution history is remembered for every process, access permissions are defined per the domain to which it belongs, and access control is performed accordingly. Figure~\ref{fig:fig-domain} is a copy of the screen of the TOMOYO Linux policy editor on Fedora~15. Each line corresponds to a different domain, and the program names of the applications executed so far are displayed. The indentation between lines indicates the parent--child relationship of domains; domains with the same indentation were created from processes of a common domain.

\subsection{Judging Application Processing}
Mandatory access control without omission must be performed in the kernel. Because the kernel's role is to process the fragmentary requests from applications, it cannot judge the intent of an application's requests or the relationships among multiple requests. The one who knows the application's intent and the relationships among its requests is the application's author---the application program itself. Even if we made an application convey its intent to the kernel, considering the case where the application is hijacked, that declaration cannot be trusted. Moreover, doing so would have the drawback of requiring modification and recompilation of existing programs. However, information that already exists within the kernel (information managed by the kernel) need not be declared by the application; it therefore requires no modification of existing programs, is unaffected by application hijacking, and can be trusted.

Within the Linux kernel there is information about processes, about i-nodes (the entities of files), about application execution requests, about files, and so on. In the specification of the policy language, we defined variables corresponding to that information, and made it possible, in the syntax of the policy language, to write kernel-internal parameters as conditional expressions, i.e., as conditions for access control.

\subsection{Implementation of MAC}
To perform access control that applies without omission and cannot be bypassed, it suffices to implement a reference monitor~\cite{sandhu1994access}\cite{Eiraku2002}. Specifically, the invocation of a system call is intercepted (hooked), and a decision is made about whether to permit the request. If permitted, the original processing is carried out; otherwise, the request is rejected and an error is returned to the caller. A reference monitor is implemented by rewriting the hooked functions, but Linux version~2.6 and later provides Linux Security Modules~\cite{Wright2003} (hereafter abbreviated LSM), a framework for extending the contents of access control processing in the kernel. Using LSM, one can hook a system call and add defined processing (here, access control) without directly rewriting the system call's processing. LSM manages the contents of system-call hooks in units called ``modules'' (also called security servers), but the current LSM specification does not allow multiple modules to be used together. In order to make coexistence with other modules possible, TOMOYO Linux provides both a version that uses LSM and a version that hooks system calls independently.

\subsection{Policy Specification}
This section shows, through concrete policy examples, that TOMOYO Linux makes possible access control taking into account a variety of conditions that conventional access control methods could not express. TOMOYO Linux supports access control not only for files but also for networks, signal transmission, and so on; but for reasons of space, we explain only access permissions related to files.

An access permission has the structure ``category\quad operation\quad mandatory-condition\quad optional-condition.'' The ``category'' is the classification of access control; the category of file-related access permissions is \texttt{file}. ``category\quad operation\quad mandatory-condition'' is mandatory. The ``operation'' is the operation about to be performed on the object; for files there are \texttt{rename}, \texttt{execute}, and so on. The ``mandatory condition'' is, for example, the path name of the file that is the target of the operation. Some path names are not determined in advance, such as temporary files created under \texttt{/tmp} with names that include a process ID. There are also cases where one wants to permit all files under a specific directory collectively, such as web content, without enumerating them. For such cases, TOMOYO Linux uses wild cards. Table~\ref{tab:tomoyo-wild} lists the wild cards available in TOMOYO Linux.

Regarding ``conditions,'' there are two kinds: mandatory conditions and optional conditions. A mandatory condition must always be specified corresponding to the contents of the ``operation,'' and the required condition differs per operation. For example, to permit \texttt{file rename}, the ``name before renaming'' and the ``name after renaming'' must always be specified, and these two are mandatory conditions.

An optional condition is used to specify, as needed, conditions that more strictly limit the granting of access; for every access permission, an arbitrary number of optional conditions can be specified.

\begin{table}[t]
\caption{TOMOYO Linux wild card patterns.}
\label{tab:tomoyo-wild}
\centering
\begin{tabular}{l|l}\hline
Pattern & Meaning\\\hline\hline
\verb|\*| & zero or more repetitions of characters other than / \\
\verb|\@| & zero or more repetitions of characters other than / and . \\
\verb|\?| & one character other than / \\
\verb|\$| & a decimal number of one or more digits \\
\verb|\+| & one decimal digit \\
\verb|\X| & a hexadecimal number of one or more digits \\
\verb|\x| & one hexadecimal digit \\
\verb|\A| & one or more alphabetic characters \\
\verb|\a| & one alphabetic character \\
\verb|\-| & operator that excludes a path name \\
\verb|/\{dir\}/| & one or more repetitions of dir/ \\\hline
\end{tabular}
\end{table}

We have explained an overview of the TOMOYO Linux policy. Details of the policy specification are published on the project web page.\footnote{\url{http://tomoyo.sourceforge.jp/1.8/policy-specification/index.html}}

\subsubsection{Example of a Domain Policy}
We give an example of the policy needed for a user who has logged in from the console to change their own password in TOMOYO Linux. In this example, first, in the domain of the shell process running after logging in from the console (\texttt{<kernel> /sbin/init /sbin/agetty /bin/login /bin/bash}), execution of the application \verb|/usr/bin/passwd| (the command that changes the password) is permitted. The resulting \verb|/usr/bin/passwd| process runs in a new domain (\texttt{<kernel> /sbin/init /sbin/agetty /bin/login /bin/bash /usr/bin/passwd}), so for that domain one must define the access permissions that \verb|/usr/bin/passwd| needs to perform the requested processing.

A TOMOYO Linux policy has a structure in which multiple per-domain definition blocks appear consecutively. Each domain block begins with the declaration of the domain name, followed by an enumeration of as many access permissions as needed for that domain.

\begin{figure}[t]
{\footnotesize
\begin{verbatim}
 1  <kernel> /sbin/init /sbin/agetty /bin/login /bin/bash
 2
 3  file execute /usr/bin/passwd exec.realpath="/usr/bin/passwd" exec.argv[0]="passwd"
 4  file read/write /dev/tty
 5  file read /etc/passwd
 6  file read /etc/profile
 7  file read /home/harada/.bash_profile
 8  file read /home/harada/.bashrc
 9  file read /etc/bashrc
10  file write /dev/null
\end{verbatim}
}
\caption{Policy of /bin/bash Domain.}
\label{fig:pol1}
\end{figure}
Line~1 of Figure~\ref{fig:pol1} indicates that the access permission definitions that follow concern a \verb|/bin/bash| whose program execution history is: executed from \verb|/sbin/init|, then \verb|/sbin/agetty|, then \verb|/bin/login|. This means the access permission definitions for the \verb|passwd| command in a shell logged in from the console. Line~3 permits the execution of an application whose path name after resolving symbolic links is \verb|/usr/bin/passwd|, on the condition that its invocation name (\texttt{exec.argv[0]}) is \verb|passwd|. If one wishes to permit execution only when specific arguments are given, one adds all the elements as conditions using \verb|exec.argv[]|. To restrict by the number of arguments, one writes, for example, \verb|exec.argc=1|.

Because TOMOYO Linux adopts the whitelist scheme (everything not explicitly permitted is denied), with this definition the only thing a shell logged in from the console may execute is the \verb|/usr/bin/passwd| satisfying the conditions written here; attempting to execute any other application results in an error. Lines~4 through 10 permit the files that the \verb|/bin/bash| belonging to this domain may read and write.

\begin{figure}[t]
{\footnotesize
\begin{verbatim}
 1  <kernel> /sbin/init /sbin/agetty /bin/login /bin/bash /usr/bin/passwd
 2
 3  file read /etc/passwd
 4  file read /etc/shadow
 5  file write /etc/.pwd.lock
 6  file read /dev/urandom
 7  file create /etc/nshadow 0666
 8  file write /etc/nshadow
 9  file chown/chgrp /etc/nshadow 0
10  file chmod /etc/nshadow 00
11  file rename /etc/nshadow /etc/shadow
\end{verbatim}
}
\caption{Policy of /usr/bin/passwd Domain.}
\label{fig:pol2}
\end{figure}
Figure~\ref{fig:pol2} gives the access permission definitions for the \verb|/usr/bin/passwd| permitted to execute in Figure~\ref{fig:pol1}; that is, for the \verb|/usr/bin/passwd| whose program execution history is ``executed from \verb|/sbin/init|, then \verb|/sbin/agetty|, then \verb|/bin/login|, then \verb|/bin/bash|.'' The \verb|passwd| command reads the contents of \verb|/etc/shadow|, writes the result of changing the password to a file called \verb|/etc/nshadow|, and then renames it to \verb|/etc/shadow|. In lines~7, 9, and 10, not only the target (object) file name but also its permission, user ID, group ID, and so on are specified as mandatory conditions; if these do not match, the access request is rejected. Because label-based MAC does not handle path names, modes, and the like, such conditions cannot be attached.

\subsubsection{Examples of Conditional Access Permission}
We show usage examples of conditional access permission, for access permissions with mandatory conditions only and for those including optional conditions.
\begin{enumerate}
\item \textbf{Examples of access permission with mandatory conditions only}\\
We show usage examples of access permission with mandatory conditions only.
\begin{itemize}
\item \texttt{file rename /etc/mtab.tmp /etc/mtab}\\Permits renaming the path name \verb|/etc/mtab.tmp| to the path name \verb|/etc/mtab|.
\item \texttt{file create /var/lock/subsys/crond 0644}\\Permits creation of \verb|/var/lock/subsys/crond| only when the permission value is 0644.
\item \texttt{file chmod /dev/mem 0644}\\Permits setting the permission of \verb|/dev/mem| only to 0644.
\item \texttt{file execute /bin/ls}\\Permits execution of \verb|/bin/ls|.
\end{itemize}
\item \textbf{Examples of access permission with optional conditions}\\
For optional conditions, the conditional expression is written in the form ``attribute=value'' (permit if the attribute equals the specified value) or ``attribute!=value'' (permit if the attribute does not equal the specified value). Multiple conditional expressions can be written, in which case access is permitted only when all of them hold. We show usage examples of access permission with optional conditions.
\begin{itemize}
\item \texttt{file symlink /dev/cdrom symlink.target="hdc"}\\Permits creating the symbolic link \verb|/dev/cdrom| only when the link's contents are \verb|hdc|.
\item \texttt{file execute /bin/bash task.uid=500-1000}\\Permits execution of \verb|/bin/bash| only when the user ID is in the range 500 to 1000.
\item \texttt{file read /tmp/file001.tmp task.uid=path1.uid}\\Permits referencing the file \verb|/tmp/file001.tmp| only when the current process's user ID matches the owner ID of that file.
\item \texttt{file execute /usr/bin/ssh exec.realpath="/usr/bin/ssh" exec.argv[0]="ssh"}\\Permits execution of \verb|/usr/bin/ssh| when the invocation name is \verb|ssh| and the path name after resolving symbolic links is \verb|/usr/bin/ssh|.
\item \texttt{file execute /bin/bash exec.realpath="/bin/bash" exec.argv[0]="-bash" task.uid!=0 task.euid!=0}\\Permits execution of \verb|/bin/bash| when the invocation name is \verb|-bash| (a login shell), the path name after resolving symbolic links is \verb|/bin/bash|, and the process's user ID and effective user ID are not 0 (the root user).
\end{itemize}
\end{enumerate}

\subsection{Operating the Policy}
This section describes the procedures for formulating and maintaining the policy needed to perform access control with TOMOYO Linux.
\subsubsection{How to Edit the Policy}
At system startup, the contents of \verb|/etc/ccs/domain_policy.conf| are read. This file is owned by the root user and can be edited with a text editor such as emacs. Access requests to this file itself are also protected according to the TOMOYO Linux policy, so after defining a domain to which access is permitted---such as ``the domain of emacs executed from a shell obtained by logging in from the console''---one defines the access permissions for operations such as read/write or read-only, adding conditions as needed.

TOMOYO Linux provides a CUI (Character User Interface) policy editor as a tool for adjusting the policy after system startup. Figure~\ref{fig:fig-domain} is the policy editor's screen; when launched, it first displays the current domain transitions. One checks the contents of the domain transitions using the cursor keys, and pressing the Return key displays the list of access permissions for the selected domain, where additions and deletions can be made. Operations in the policy editor are immediately reflected in the contents of access control. Other policy-editing tools are provided for loading and saving policies. For the contents and usage of the management tools provided for TOMOYO Linux, please refer to the project web page.\footnote{\url{http://tomoyo.sourceforge.jp/1.8/man-pages/index.html}}

\subsubsection{Flow of Policy Formulation}
We describe the rough flow of policy formulation in TOMOYO Linux. For more detailed contents, please refer to the tutorial published by the project.\footnote{\url{http://tomoyo.sourceforge.jp/about.html}}

TOMOYO Linux has the concept of a control mode. The control mode specifies the operational behavior of access control; there are four kinds: disabled, learning, permissive, and enforcing. The meaning of each mode is shown in Table~\ref{tab:mode}.

\begin{table}[t]
\caption{TOMOYO Linux mode.}
\label{tab:mode}
\centering
\begin{tabular}{l|l|p{0.5\linewidth}}\hline
Mode & Meaning & Behavior\\\hline\hline
disabled & disabled & Operates the same as an ordinary kernel\\
learning & learning & Does not reject a request even if a policy violation occurs; adds to the policy the access permissions needed so that policy violations do not occur\\
permissive & confirmation & Does not reject a request even if a policy violation occurs\\
enforcing & enforcement & Rejects a request if a policy violation occurs\\\hline
\end{tabular}
\end{table}

The mode can be specified per domain, either defined in the policy file or changed in the policy editor. The basic flow of policy formulation in TOMOYO Linux is as follows.

\begin{enumerate}
\item Operate the system in learning mode for a certain period. This collects the program execution histories (the list of domains) and the access-request information of each domain.
\item Check the results obtained in the previous step and create the policy definitions.
\item Operate the system in permissive mode and check whether any access permissions are missing.
\item Operate the system in enforcing mode.
\end{enumerate}

The period required for each step varies with the contents of the system and what one wishes to restrict. The system administrator must judge the timing for operating in enforcing mode (the timing for freezing the policy). With the proposed method, by executing the needed applications in the needed states (domains), one can obtain the execution histories. Because the contents automatically include the group of applications invoked from within the application in question (and further the groups of applications invoked from those), executing all related applications in every situation of the current system yields the complete execution histories needed. Therefore, when this is feasible, it suffices. In general, however, the number of combinations is enormous, and exhausting all of them is difficult. This problem is analogous to the common practice of, in order to remove bugs from a system, executing every pattern and checking the results. Therefore, like that practice, we believe it appropriate in practice to prevent execution omissions with the following approach, and we actually operate in this way.

\begin{itemize}
\item Collect program execution history information over as long a period as possible.\\
By lengthening the observation period, the possibility of application execution omissions can be reduced. In the demonstration experiment of applying TOMOYO Linux to a web server described later, this was done over about two weeks.
\item Use evaluation programs, test tools, and the like if available.\\
If there are evaluation programs or test tools prepared for each application, executing them allows the applications to be executed comprehensively.
\item Test thoroughly.\\
The testing referred to here is not testing of the policy, but of the computer to be protected itself and of the software that provides services on it.
\item Check batch processing and error handling.\\
Even after doing all of the above, there is a possibility of omissions for applications that are not always running, such as monthly batches and disaster recovery. Batch processing can be checked by changing the system time; disaster recovery and the like can be checked by artificially causing the situations that trigger them.
\end{itemize}

Once the contents have been settled, the system is started with the protective function rather than the confirmation function.

\subsubsection{Updating the Policy}
By switching the control mode to enforcing, protection by TOMOYO Linux becomes effective. The system administrator checks, via the log, for the presence and contents of access requests not in the policy. If, after checking the contents of the requests left in the log, the administrator judges that a given request should be permitted, the administrator adds the corresponding contents in the policy editor and grants permission. Because the contents of the TOMOYO Linux log include the domain name, the situation in which the error occurred can be identified, and the place where it should be reflected is clear.

For OS updates and the addition or removal of applications (packages), if their effects are known in advance, one can respond by modifying the policy in the policy editor or by editing \verb|/etc/ccs/domain_policy.conf| with a text editor. Otherwise, one temporarily disables MAC, performs the update, installation, or removal, and then modifies the policy. This is an operation common to MAC, not limited to TOMOYO Linux. The reason for disabling MAC here is that, because MAC decides access requests regardless of administrative privileges, it is conceivable that updates and the like would not be applied.

\section{Evaluation}\label{sec:Consideration}
This chapter, as an evaluation of the proposed method, organizes its features and describes the contents of an initial study of its usability. It also shows the results of measuring, using TOMOYO Linux, the processing delay caused by introducing the proposed method.

\subsection{Features of the Proposed Method}
The proposed method of this research has the following features.
\begin{enumerate}
\item \textbf{Access control taking application processing into account can be performed}\\
As shown in the examples of the previous chapter, access control can be performed that takes into account an application's processing and the state of the system after access is permitted.
\item \textbf{The system's behavior can be understood and configured}\\
In a system that has introduced MAC, one monitors for access requests not in the policy and, when one occurs, distinguishes whether it is an omission in the policy definition or an illegitimate request, responding accordingly. With the proposed method, when an access request results in an error, the program execution history information is obtained, so confirming the situation is easy.
\item \textbf{Access permissions can be defined independently per application (domain)}\\
In SELinux, an implementation of label-based MAC, the necessary labels are defined for the entire information system as the target. When an error occurs in an application with a certain label, one should respond after considering the group of applications with that label and the transition conditions to other labels, which is not necessarily easy. As seen in the policy examples shown in Chapter~4, the policy of this access control method is defined independently per domain and can be modified, so management is easy.
\item \textbf{It can be used as Role-Based Access Control}\\
The access control method proposed in this paper can also be used in the manner of the Role-Based Access Control Model~\cite{Sandhu1996} (RBAC) or the Identity-Based Access Control Model. As shown in the example of a shell's program execution history in Chapter~3, the shell obtained by executing \verb|/bin/su| from a login shell can have its access permissions defined independently of the original shell. The same holds for nesting shells beyond \verb|/bin/su|; by changing the access permissions granted according to the shell hierarchy, it is also possible to assign roles per application executed from a shell.~\cite{Harada2005a} By writing user IDs and group IDs as conditions, conditions for permitting access---root only (the system administrator account in Linux/UNIX), non-root, a specific user ID, group ID, and so on---can be written flexibly.
\end{enumerate}

\subsection{Initial Study of Usability}
Regarding the operation of the proposed method, we performed a desk evaluation against SELinux, which adopts label-based MAC. We also introduce the results of a demonstration experiment conducted on TOMOYO Linux in fiscal year 2007.

\subsubsection{Effects on Operation}
Based on the results of a survey using test subjects, the Information-technology Promotion Agency (an independent administrative institution) published a ``Guideline for Building a Secure Web Server'' using SELinux~\cite{ipa2003}. That guideline concludes that, for applying SELinux, a large amount of work is required for surveying the specifications of the target applications and for surveying, analyzing, and modifying the ``standard security policy''\footnote{This refers to a policy matched to the standard settings of the aforementioned distribution.} definitions, or for defining a new security policy. The provision of a standard security policy is a boon for system administrators, but the environment assumed by the standard security policy does not match the environment of the system that the administrator manages. To match the contents of the standard security policy to the system being managed, or to consider how to respond when an error occurs, the system administrator must understand an enormous policy.

With the proposed method, the classification of an application's execution state is performed automatically within the kernel, and the relationships among classifications are easy to understand, so the system administrator can concentrate on the access permission definitions for each execution state (domain).

\subsubsection{Evaluation of Introduction on a Real Server}
In fiscal year 2007, the Secure OS Promotion WG (at the time) of the NPO Japan Network Security Association conducted a demonstration experiment of introducing TOMOYO Linux on the association's own web server, with the aim of evaluating the contents and effects of introducing a secure OS and its impact on system management. In the experiment, file-related access control was configured for an Apache web server published on the Internet. The web server used CGI, but the CGI was managed as a domain separate from Apache, and from the history of file-access requests collected via the log, the access permissions needed for each domain were defined on a per-file basis. Participants in the experiment were able to understand the operating principles of TOMOYO Linux and the editing of the policy in a very short time, and there were comments that, through the work of formulating the policy, they were able to grasp the behavior of the Linux server. These effects are thought to be due to the fact that path names can be written directly in the policy, and that the classification of access control by program execution history is easy for administrators to understand. A report by the participating members has been published.\footnote{\url{http://www.jnsa.org/result/2007/tech/secos/}}\footnote{Excerpts of the policy are also included, but they differ from the examples in this paper due to version differences.}

\subsection{Impact on Performance}
\label{bench}
This section shows and discusses the results of measuring, using TOMOYO Linux, the impact of the proposed method on performance---the impact by system-call type and the impact by policy scale.

\subsubsection{Impact by System-Call Type}
Using LMBench~\cite{mcvoy1996lmbench}, a benchmark tool for UNIX-family systems, we measured the impact by system-call type.

When the proposed method is applied, in principle the performance impact applies to hooked system calls, and system calls that are not hooked are expected to be unaffected. Therefore, among the OS-related system calls measurable by LMBench, we compare those that TOMOYO Linux does not hook and those that it does hook separately. As for the comparison conditions, we ran LMBench in a state where TOMOYO Linux was not installed and in a state where access control by TOMOYO Linux was enabled.

The execution environment of LMBench is shown in Table~\ref{tab:lmbenchenv}. The measurement items and detailed specifications available in LMBench are posted on that tool's web page,\footnote{\url{http://www.bitmover.com/lmbench/}} so we do not explain them here. The measurement results for system calls that TOMOYO Linux does not hook are shown in Table~\ref{tab:lmbench-no-hook}, and the results for system calls that TOMOYO Linux does hook are shown in Table~\ref{tab:lmbench-hook}. The benchmarks were conducted after ensuring that no processes other than those needed to run the benchmark were running. We explain how to read Tables~\ref{tab:lmbench-no-hook} and \ref{tab:lmbench-hook}. The column Func.\ is the LMBench test item; Base is the measurement result ($\mu$sec) in the state where TOMOYO Linux is not installed; TOMOYO is the measurement result ($\mu$sec) in the state where TOMOYO Linux is introduced and MAC is enabled; Diff is the difference ($\mu$sec) between TOMOYO and Base; and Overhead is the value obtained by
\[ \mathit{Overhead} = \frac{\mathit{TOMOYO} - \mathit{Base}}{\mathit{Base}} \times 100. \]
An Overhead value of 100 means that a 100\% delay occurred relative to the case where TOMOYO Linux is not enabled---that is, the time required for processing doubled.

\begin{table}[t]
\caption{Benchmark Environment.}
\label{tab:lmbenchenv}
\centering
\begin{tabular}{l|l}\hline
& specification/version\\\hline\hline
CPU & Core 2 Duo T7200 2.0GHz\\
Memory & 2GB\\
OS & Ubuntu 10.04 x86\_64\\
Kernel & 2.6.32-39.86\\
TOMOYO Linux & 1.8.3p5\\
Benchmark tool & LMBench 3.0-a9\\\hline
\end{tabular}
\end{table}

\begin{table}[t]
\caption{Result of LMBench (not hooked).}
\label{tab:lmbench-no-hook}
\centering
\begin{tabular}{l|r|r|r|r}\hline
Func. & Base ($\mu$sec) & TOMOYO ($\mu$sec) & Diff ($\mu$sec) & Overhead (\%)\\\hline\hline
null syscall & 0.274 & 0.269 & 0.0 & -1.82\\
null I/O & 0.4365 & 0.418 & 0.0 & -4.24\\
Select on 100 tcp fd's & 7.0815 & 7.1455 & 0.1 & 0.90\\
Signal handler installation & 0.552 & 0.56 & 0.0 & 1.45\\
2p/0K ctxsw & 10.97 & 10.665 & -0.3 & -2.78\\
2p/16K ctxsw & 11.26 & 11.07 & -0.2 & -1.69\\
2p/64K ctxsw & 14.21 & 14.39 & 0.2 & 1.27\\
8p/16K ctxsw & 12.22 & 11.755 & -0.5 & -3.81\\
8p/64K ctxsw & 14.035 & 14.095 & 0.1 & 0.43\\
16p/16K ctxsw & 12.185 & 12.04 & -0.1 & -1.19\\
16p/64K ctxsw & 14.135 & 14.225 & 0.1 & 0.64\\
Pipe & 38.3 & 36.73 & -1.6 & -4.10\\
AF UNIX & 24.47 & 24.28 & -0.2 & -0.78\\
Mmap & 2341.55 & 2375.1 & 33.5 & 1.43\\
Page Fault & 2.52829 & 2.58089 & 0.1 & 2.08\\
Select on 100 fd's & 3.254 & 3.35045 & 0.1 & 2.96\\\hline
\end{tabular}
\end{table}

\begin{table}[t]
\caption{Result of LMBench (hooked by TOMOYO).}
\label{tab:lmbench-hook}
\centering
\begin{tabular}{l|r|r|r|r}\hline
Func. & Base ($\mu$sec) & TOMOYO ($\mu$sec) & Diff ($\mu$sec) & Overhead (\%)\\\hline\hline
Simple stat & 3.12 & 7.1145 & 4.0 & 128.03\\
Simple open/close & 5.037 & 9.5065 & 4.5 & 88.73\\
Signal handler overhead & 3.8015 & 5.961 & 2.2 & 56.81\\
Process fork+exit & 300.35 & 301.7 & 1.3 & 0.45\\
Process fork+execve & 1001.95 & 1062.55 & 60.6 & 6.05\\
Process fork+/bin/sh -c & 2226.1 & 2551.2 & 325.1 & 14.60\\
UDP & 54.65 & 69.91 & 15.3 & 27.92\\
RPC/UDP & 61.565 & 80.615 & 19.1 & 30.94\\
TCP & 58.04 & 57.3 & -0.7 & -1.27\\
RPC/TCP & 72.64 & 71.36 & -1.3 & -1.76\\
TCP/IP connection cost & 65.8 & 72.4 & 6.6 & 10.03\\
0K File Create & 25.32 & 41.035 & 15.7 & 62.07\\
0K File Delete & 19.865 & 27.795 & 7.9 & 39.92\\
10K File Create & 80.9 & 95.855 & 15.0 & 18.49\\
10K File Delete & 41.075 & 50.33 & 9.3 & 22.53\\\hline
\end{tabular}
\end{table}

Looking at Table~\ref{tab:lmbench-no-hook}, the proportion of delay when TOMOYO Linux is enabled is within $\pm5\%$. In theory, enabling TOMOYO Linux cannot improve processing speed, so these are considered to be measurement errors of LMBench.

Looking at Table~\ref{tab:lmbench-hook}, for \texttt{stat}, \texttt{open/close}, and \texttt{signal handler}, a delay of more than 50\% occurs. For \texttt{0K File Create} (creating an empty file), the delay is more than 60\%, but for \texttt{10K File Create} the value is conversely smaller, 18.49\%. The latter, after creating an empty file, performs a \texttt{write} of 10KB of data; because processing not hooked by TOMOYO is added, the relative proportion of the delay time becomes smaller, which is considered to be why it appears this way.

For the \texttt{fork}-related items, which are considered to be affected by program-execution-history processing, a delay of about 5\% occurs for the combination of \texttt{fork} and \texttt{exec}. The value for \texttt{fork+/bin/sh -c} is larger, which is considered to be because the \texttt{exec} of \texttt{/bin/sh} is performed, so the \texttt{exec} processing is executed twice.

As an attempt to evaluate the performance of MAC (secure OS) using LSM, there is the LSM Performance Monitor (LSMPMON)~\cite{Matsuda2009}. LSMPMON can record the number of calls and the processing time of hook functions, but in terms of predicting the impact on performance as perceived by users and system administrators, the problem has not been solved, and future research is anticipated.

\subsubsection{Impact by Policy Scale}
We evaluated the impact on performance caused by an increase in the number of applications executed---and a consequent increase in policy definitions---due to package installation and the like.

In the proposed method, at application execution (when the \texttt{execve} system call is executed), the domain corresponding to that process is assigned. Next, a judgment is made as to whether the requested content is included in the access permissions of that domain. As the number of applications executed increases and the number of domains increases, the processing time to search for the corresponding domain per access request increases. Also, as application processing becomes more complex, the number of access permissions per domain increases, and the processing time to search for the corresponding access permission increases. Therefore, the impact on performance caused by an increase in policy scale can be measured by the following two.

\begin{itemize}
\item The increase in domain-search time due to an increase in the number of domains
\item The increase in access-permission-search time due to an increase in the number of access permissions
\end{itemize}

\begin{figure}[t]
 \centering
  \includegraphics[width=.7\linewidth]{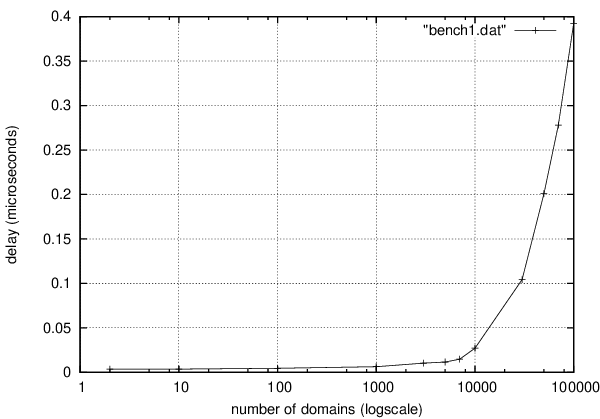}
 \caption{Performance delay due to domain number increase.}
 \label{fig:fig-bench1}
\end{figure}

For the increase in domain-search time, we created a program that mechanically generates a specified number of domains, and measured the processing time when executing \texttt{/tmp/reexec} (a program that executes itself a specified number of times) for a number of domains ranging from 2 to 100000. For each case, the domain for \texttt{/tmp/reexec} is made to be found last. The relationship between the number of domains and the processing time is shown in Figure~\ref{fig:fig-bench1} (the number of domains uses a logarithmic axis).

For the case where the number of domains is 2, the difference between when TOMOYO Linux is enabled and when TOMOYO Linux is not installed was 0.00362~$\mu$sec. In Figure~\ref{fig:fig-bench1}, the number of domains uses a logarithmic axis, and as expected the delay in the time required for domain search increases with the number of domains; we see that the rate of increase becomes pronounced once the number of domains exceeds 10000.

\begin{figure}[t]
 \centering
  \includegraphics[width=.7\linewidth]{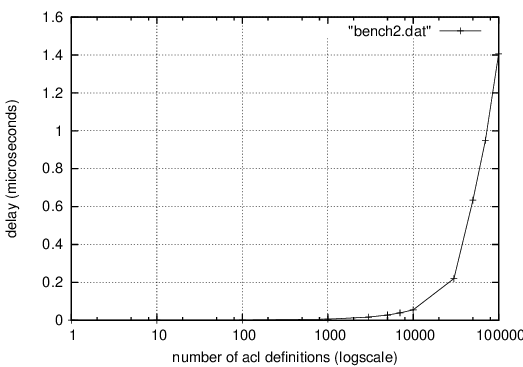}
 \caption{Performance delay due to ACL number increase.}
 \label{fig:fig-bench2}
\end{figure}

For the delay in access-permission-search time, we created a program that mechanically generates a specified number of access permissions, and measured the time to \texttt{open} \verb|/dev/null| 10000 times for a number ranging from 1 to 100000. For each case, the entry permitting the \texttt{open} of \verb|/dev/null| is made to be found last. The relationship between the number of access-permission lines and the processing time is shown in Figure~\ref{fig:fig-bench2}. The number of access-permission lines uses a logarithmic axis. For the case where the number of access permissions is 1, the difference between when TOMOYO Linux is enabled and when TOMOYO Linux is not installed was 0.0032~$\mu$sec.

In Figure~\ref{fig:fig-bench2}, the number of access permissions uses a logarithmic axis, and as expected the time required to search access permissions increases with the number of access permissions; we see that the rate of increase becomes pronounced once the number of access permissions exceeds 10000.

As an empirical rule from the first release in 2005 up to the present, it has been confirmed that when TOMOYO Linux is introduced and used on a major Linux distribution, the number of domains and the number of access permissions per domain each fall within a range of less than 2000. Adding packages and increasing the number of applications executed increases the number of domains, but looking at the results in Figures~\ref{fig:fig-bench1} and \ref{fig:fig-bench2}, it is considered that no major performance hindrance arises in ordinary use.

\section{Discussion}
This chapter, as a discussion of the proposed method, compares it with other closely related methods and confirms its effect against typical unauthorized-access techniques. It also describes cases that the proposed method cannot solve and, as countermeasures, the issues in enhancing its functions.

\subsection{Comparison with Other Methods}
\subsubsection{Comparison with Label-Based MAC}
\begin{enumerate}
\item \textbf{Features and suitable uses of label-based MAC}\\
In label-based MAC, access decisions for subjects and objects can be made without being affected by name changes such as renaming. Also, because it does not use names, it can be applied even to resources that have no name. For these reasons, label-based MAC can be said to be suited to the use of robustly restricting whether a subject may access an object.
\item \textbf{Features and suitable uses of the proposed method}\\
In the proposed method, by utilizing the factors that affect an application's processing as parameters for access decisions, the cases in which access is permitted can be narrowed down. Because file names, DAC permissions, and the like can be used as parameters, their changes can be kept within the range that the system administrator has permitted. For this reason, the proposed method can be said to be suited to the use of correctly maintaining the state of an information system and not allowing extraneous functions to be used.
\item \textbf{Positioning of the proposed method}\\
Which method is suitable depends on what threat one wishes to prevent. At the time the TCSEC was published, in 1983, the leakage of confidential information was the chief concern and the threat to be prevented. It was important that only qualified applications could access the files they needed, and that information obtained from files did not leak. Label-based MAC satisfies this need, but no attention was paid to the contents of an application's processing. The proposed method focuses on the contents of the processing performed by each application's process and aims to prevent it from performing anything other than the necessary processing. The proposed method can be positioned not as a replacement for label-based access control methods but as a complement to their limits.
\end{enumerate}

\subsubsection{Comparison with AppArmor}
TOMOYO Linux was released in November 2005, and AppArmor in January 2006. Because, until they were released, neither knew that the other was developing pathname-based MAC, they have different implementation schemes despite both being pathname-based.
\begin{enumerate}
\item \textbf{Differences in implementation}\\
TOMOYO Linux, before AppArmor, recognized the need for access decisions that take various factors (condition parameters such as command-line arguments and environment variables) into account, actually implemented it, and it has already been adopted into the mainline. In AppArmor, discussion of which conditions (parameters) should be checked has also begun, and in November 2011 a proposal to use the values of environment variables in access control was made on the AppArmor developers' mailing list.\footnote{\url{https://lists.ubuntu.com/archives/apparmor/2011-November/001668.html}} In the mailing-list discussion, the specification and implementation method of TOMOYO Linux were introduced, and there is a possibility that they will be adopted in AppArmor in the future.
\item \textbf{Applicability of the proposed method}\\
AppArmor and TOMOYO Linux have in common that they use path names for objects, but their implementation schemes for classifying subjects differ greatly, and it is not appropriate to regard them as the same method. Specifically, because TOMOYO Linux records the entire history from startup and can define the policy based on its contents, it can realize access restriction targeting all processes from startup to shutdown. Because AppArmor does not remember the history from startup, it can realize access control targeting only specific processes such as daemons and web browsers. This is a difference originating from a difference in design philosophy.
\item \textbf{Ingenuities of TOMOYO Linux not found in AppArmor}\\
The ingenuities of TOMOYO Linux not found in AppArmor are shown below.
\begin{itemize}
\item Access restriction targeting all processes from system startup to shutdown is possible.
\item By applying the fact that domain transitions form a tree structure, RBAC-like usage and the strengthening of login authentication are also possible.
\item Because the scheme escapes characters interpreted as wild cards with \textbackslash{} (backslash), the kinds of wild cards can be increased while maintaining compatibility.
\item To exclude directory names or file names with a specific meaning, such as \texttt{.git}, it has a subtraction operator \textbackslash\texttt{-}.
\item Because it does not use a ready-made policy, a policy with no waste (no unnecessary access permissions) can be defined for the target system. And because one configures it while understanding the contents oneself, one does not end up disabling it when a problem arises.
\end{itemize}
\end{enumerate}

\subsubsection{Comparison with Context-Aware Access Control Efforts}
In recent years, efforts related to access control that reflects context (CAAC: Context-Aware Access Control) have been actively pursued.

In ``A survey on context-aware systems'' by Matthias Baldauf et al.~\cite{baldauf2007survey}, a context-aware system is defined as ``one that aims to improve usability and efficiency by taking environmental context into account without requiring active intervention by the user.'' Today, many efforts related to CAAC treat location information and network information as context in the mobile and ubiquitous domains, or attempt to introduce the concept of context into web services~\cite{truong2009survey}. Context in CAAC is an abstract concept; if one understands it as aiming to create added value by gathering and utilizing available information, then the method proposed in this paper can also be classified as related to CAAC.

Among efforts related to CAAC, one close to the proposal of this paper is Salvia~\cite{Suzuki2006}. Salvia divides context into two kinds, internal and external, and has built a prototype that uses, as internal context, the attribute values of processes obtainable inside the OS and the history of system calls, and, as external context, location information, absolute time, wireless LAN access points (ESSID), and so on. In its use of process attribute values as internal context, it is close to the idea of judging application processing in this paper, and in that the system-call history aims at finer-grained subdivision of process state, its aim is close to the proposal of this paper. However, to infer the situation in which a process is executed from the system-call number, arguments, and return value, detailed analysis is required. Extracting only specific cases is possible, but using it comprehensively is considered difficult. The proposed method has the advantage that it can be realized more simply by using the domain as internal context.

\subsection{Effect Against Unauthorized-Access Techniques}
We evaluate the effect of the proposed method against typical unauthorized-access (attack) techniques~\cite{Shinagawa2004}.

\begin{enumerate}
\item \textbf{Execution of unauthorized applications}\\
Here, ``unauthorized'' refers to the execution of applications the system administrator does not desire. A typical example is the case of exploiting a vulnerability in a server providing a service over a network to take it over and execute a shell (and the applications executed from that shell). Buffer overflow attacks, format string attacks, OS command injection attacks, and the like are similar. In a buffer overflow attack, it is common to launch a shell and then perform unauthorized operations. If, for the target application, execution of \verb|/bin/sh| is not permitted, then even if a vulnerable application is seized, the launching of a shell fails, and subsequent operations can be prevented. To do this with label-based MAC, label definitions for subjects and objects are needed, taking into account the subject's situation and its transitions, and the combinations of objects for subjects placed in each situation. In the proposed method of this research, because each process has a program execution history that indicates the situation it is placed in, the administrator does not need to newly define it. Also, when execution of a shell is needed as legitimate processing, its execution must be permitted regardless of the method; but using the proposed method, command-line arguments, environment variables, and the like can be used as constraint conditions on application execution to limit it, so the risk of misuse can be reduced.
\item \textbf{Path traversal attacks}\\
There is an attack that uses relative paths and the like to read or write files that an application should not originally access. When the method proposed in this paper is implemented on Linux, by computing the absolute path name back from the kernel-internal data structures derived from the program execution history and path names, and using that path name, path traversal attacks can be nullified.
\item \textbf{Symbolic link attacks}\\
A symbolic link has the effect of making a target file (path name) appear to be a different file. The attack of exploiting this to make an application destroy the file pointed to by the link is called a symbolic link attack. When the method proposed in this paper is used, the attack does not succeed unless the creation of a symbolic link is first permitted in the directory used for the attack. Even if it is permitted, the link target of the symbolic link to be created (the string of the file name or directory name) can be limited as a constraint condition, so the risk of misuse can be reduced.
\end{enumerate}

\subsection{Issues Toward Functional Enhancement}
The proposed method, triggered by an application execution request (\verb|execve()|), derives the program execution history per access subject and classifies the subject's execution state. Therefore, unless \verb|execve()| is involved, there is a problem that the execution state of the access subject cannot be separated.

For example, the Apache web server supports CGI (Common Gateway Interface), which is commonly used; in the proposed method, when the CGI's execution method involves \verb|execve()|, that CGI becomes an independent domain, but when it does not involve \verb|execve()|, as with \verb|mod_perl|, it ends up in the same domain as Apache. For this, if a mechanism is realized by which an application conveys the information that it wants to newly separate a domain, then even CGIs that do not involve \verb|execve()| can be treated as independent domains.

The proposed method has the advantage that it can be applied to any application without requiring modification of the application; but by also using the application's own declaration, the application's author can more directly and concretely exclude unnecessary access requests. Realizing this declaration scheme is a future issue.

As another problem, for example, in cases such as permitting access to a temporary file placed in a shared directory only to the application that created it, it is necessary to remember, for each file, information about its owner and to perform access control based on that. In such cases, the issue can be realized by using the proposed method together with label-based MAC.

In addition, for an application that runs virtualized servers, applications that have the same program execution history cannot have their access-control permissions distinguished even if the contents being executed are different virtual machines. A method to restrict the range accessible per virtual machine and apply it to such cases so that they do not affect one another is also a future issue.

\section{Conclusion}\label{sec:Epilogue}
In this paper, we proposed an access control method that focuses on the contents of an application's processing---something not considered in conventional access control---and showed the method for realizing it, taking as an example TOMOYO Linux, which the authors implemented on Linux. The proposed method makes possible access control that takes into account, for each execution state of an application, the processing each application is about to perform. By this, unauthorized and unnecessary access requests that conventional access control methods could not identify can be rejected, and information security can be enhanced. We showed the effectiveness and applicability of the proposed method through examples of TOMOYO Linux policies, and introduced, regarding the usability of an information system implementing the proposed method, a desk comparison with label-based MAC and the results of the TOMOYO Linux demonstration experiment. We also discussed the effect against typical unauthorized-access techniques and confirmed that the proposed method is effective against typical attack techniques and has no important weakness originating from the method. Regarding the degree of impact on performance, we presented a discussion of the contents and degree of impact using measurement results with TOMOYO Linux. In the future, we plan to improve the accuracy of the application execution state through the combined use of declarations from applications, and to study methods for verifying usability.

\section*{Acknowledgment}
The development of TOMOYO Linux was conducted as a corporate activity of NTT DATA CORPORATION.
We would like to express our gratitude to Mr.~Shinichi Yamada, Mr.~Kazuo Tanaka, and others who have understood and supported this initiative over a long period.
We pray for the repose of the soul of the late Mr.~Yukio Itakura, who enthusiastically guided and encouraged this research, and we express our deepest appreciation for his mentorship.

\bibliographystyle{unsrtnat}
\bibliography{refs}

\section*{Author Biographies}
 
\noindent \textbf{Toshiharu Harada}\\
Graduated from the Department of Applied Physics, Faculty of Engineering, Hokkaido University in 1985. Joined Nippon Telegraph and Telephone Corporation (NTT) in the same year. Served as a visiting researcher at MIT for two years starting in 1991. After engaging in the development of multimedia authoring systems and BS/terrestrial digital data broadcasting, he has been involved in open-source project management, including thin clients and security enhancements, since 2003. Doctoral student at the Institute of Information Security. Member of IPSJ, IEICE, IEEE-CS, and ACM.
 
\vspace{0.5em}
\noindent \textbf{Tetsuo Handa}\\
Graduated from the Department of Electrical Engineering and Electronics, College of Science and Engineering, Aoyama Gakuin University in 2001. Joined NTT DATA CUSTOMER SERVICE Corporation in the same year. Currently working at NTT DATA INTELLILINK CORPORATION. Has been engaged in the research and development of TOMOYO Linux since 2003.
 
\vspace{0.5em}
\noindent \textbf{Masaki Hashimoto}\\
Graduated from the Institute of Arts and Humanities, Ritsumeikan University in 2001. Established Wipe Inc. while a student and served as the board member in charge of information systems. Received his Ph.D. in Informatics from the Institute of Information Security in 2010. Since April 2010, he has been an Assistant Professor at the Institute of Information Security, where he primarily conducts research and education on OS-based access control. Member of IPSJ, IEICE, and IEEE. Expert committee member of the IEICE Technical Committee on Information and Communication System Security (ICSS).
 
\vspace{0.5em}
\noindent \textbf{Hidehiko Tanaka}\\
Completed the doctoral course in Electrical Engineering at the Graduate School of Engineering, The University of Tokyo, in 1970. Ph.D. in Engineering. Engaged in education and research on computer architecture, parallel processing, artificial intelligence, natural language processing, distributed processing, and media processing at The University of Tokyo. After serving as Professor at the Faculty of Engineering and Dean of the Graduate School of Information Science and Technology at The University of Tokyo, he became Dean and Professor at the Institute of Information Security in 2004. Honorary Member of IPSJ. Recipient of numerous awards, including the JSAI Best Paper Award, ACM SIGGRAPH '99 Impact Paper Award, JSAI Distinguished Service Award, Tokyo Metropolitan Commendation for Science and Technology, and the Minister of Economy, Trade and Industry Award. Council Member of the Research Organization of Information and Systems, Member of the Science Council of Japan, IEEE Fellow, and Professor Emeritus at The University of Tokyo.

\end{document}